\documentclass[twoside]{article}
\usepackage{fleqn,espcrc2,psfig}
\usepackage{graphicx}
\usepackage[figuresright]{rotating}

\newcommand{\AmS}{{\protect\the\textfont2  
A\kern-.1667em\lower.5ex\hbox{M}\kern-.125emS}}
\begin{document}
\title{SDLCQ and String/Field Theory Correspondences\thanks{Based on work 
with S. Pinsky, O. Lunin, and J.R. Hiller.}}
\author{Uwe Trittmann, Department of Physics, 
	Ohio State University, Columbus, OH, USA}
%
%

\begin{abstract}
String/Field theory correspondences have been discussed heavily in recent
years. Here, we describe a testing scenario involving a non-perturbative 
field theory calculation using the framework of supersymmetric discrete 
light-cone quantization (SDLCQ).
We consider a Maldacena-type conjecture applied to the near horizon 
geometry of 
a D1-brane in the supergravity approximation. Numerical results 
of a test of this conjecture are presented with orders of magnitude more
states than we previously considered.
These results support the Maldacena conjecture and are within 10-15\% 
of the predicted results.
We present a method for using a ``flavor'' symmetry to greatly reduce 
the size of the Fock basis and discuss a numerical method that we use which is
particularly well suited for this type of matrix element calculation. 
Our results are still 
not sufficient to demonstrate convergence, and, therefore, cannot be 
considered to be a numerical proof of the conjecture. 
We update our continuous efforts to improve on these results and present
some results on the way to higher dimensional scenarios.
\vspace{1pc}
\end{abstract}
\maketitle
%
%
\def\d{\partial}
\def\beq{\begin{equation}}
\def\eeq{\end{equation}}

\section{Introduction}

The conjecture that certain field theories 
admit concrete realizations as string theories on particular backgrounds
has caused a lot of excitement in the last years. Originally, the so-called
Maldacena conjecture \cite{Maldacena} assured that the ${\cal N}=4$ 
supersymmetric Yang-Mills (SYM) theory in 3+1 dimensions is
equivalent to Type IIB string theory on an $AdS_5\times S^5$ background.
Meanwhile, other string/field theory correspondences have been conjectured.
Attempts to rigorously 
test these conjectures have met with limited
success, because our understanding of both sides of the correspondences is
usually insufficient. 
The main obstacle is that at the point of correspondence, we require two 
conditions to hold which are mutually exclusive. Namely, we want a situation 
where the curvature of the considered space-time is small in order to
be able to use the supergravity approximation to string theory.  
One also desires the corresponding field theory to be in 
a small coupling regime. So far it has been impossible
to find such a scenario.
We present a way out of this dilemma by relaxing the second requirement and
performing 
a {\em non-perturbative} calculation on the field theory side. 
To create a manageable situation, we chose a special string/field theory
correspondence in order to 
apply the non-perturbative method, namely SDLCQ, at its optimal working point.

SDLCQ, or Supersymmetric Discretized Light-Cone Quantization, is a 
non-per\-turbative method for solving bound-state problems that
has been shown to have excellent convergence properties, in particular in 
low dimensions. 
Therefore, we are looking for a (preferably) two-dimensional field theory,
which is conjectured to be equivalent to a string theory.
It turns out that the 
Yang-Mills theory with 16 supercharges in two dimensions \cite{Anton98}
has its corresponding string theory partner in  
a system of D1-branes in Type IIB string theory decoupling from gravity 
\cite{Itzhaki}. 
Since both systems have separately been studied in the literature already,
this systems is an optimal candidate to 
study the string/field theory correspondence.

The next step is to find an observable that can be computed relatively easy 
on both sides of the correspondence.
It turns out that the correlation function of a gauge invariant operator is a 
well-behaved object in this sense. 
We chose the stress-energy tensor $T^{\mu\nu}$ as this operator 
and will construct this observable in the supergravity approximation
to string theory
and perform a non-perturbative SDLCQ calculation of this correlator 
on the field theory side.

\begin{figure*}
\vspace*{0.5cm}
\centerline{
\unitlength0.9cm
\begin{picture}(15,2)\thicklines
\put(0.2,1){\vector(1,0){14.8}}
\put(0.2,0.8){\line(0,1){0.4}}
\put(5,0.8){\line(0,1){0.4}}
\put(10,0.8){\line(0,1){0.4}}
\put(0,0){$0$}
\put(4.5,0){$\frac{1}{g_{YM}\sqrt{N_c}}$}
\put(9.5,0){$\frac{\sqrt{N_c}}{g_{YM}}$}
\put(14.8,0.2){$x$}
\put(2,0.3){UV}
\put(6.5,0.3){SUGRA}
\put(12,0.3){IR}
\put(1.5,1.5){$N_c^2/x^4$}
\put(6.0,1.5){$N_c^{3/2}/(g_{YM} x^5)$}
\put(11.5,1.5){$N_c/x^4$}
\end{picture}
}
\caption{Phase diagram of two-dimensional ${\cal N}=(8,8)$ SYM:
the theory flows from a CFT in the UV to a conformal $\sigma$-model in the IR.
The SUGRA approximation is valid in the intermediate range of distances,
$1/g_{YM}\sqrt{N_c}< x < \sqrt{N_c}/g_{YM}$.
\label{figphase}}
\end{figure*}
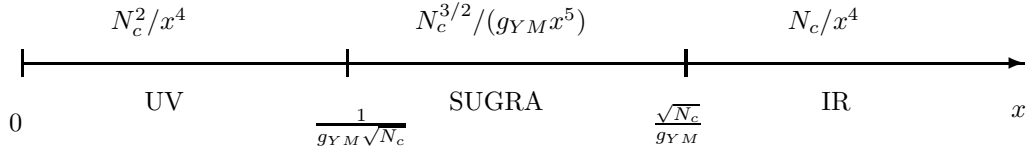

\section{Correlation functions from supergravity}

It is instructive to take a closer look on the expected properties
of ${\cal N}=(8,8)$ SYM, before we proceed to technical details on 
the string theory side. 
In the extreme ultra-violet (UV) this theory is conformally free and 
has a central charge $c_{UV}=N^2_c$. Perturbation theory in turn will be 
valid for small effective couplings $g=g_{YM}\sqrt{N_c}x$, where $x$ is
a space coordinate. For large distances, in the far infra-red (IR),
the theory becomes a conformal $\sigma$-model with target space 
$(R^8)^{N_c}/S_{N_c}$. The central charge is $c_{IR}=N_c$.
It is a bit more involved to show that here perturbation theory breaks down 
when $x{{\sim}} \sqrt{N_c}/g_{YM}$, 
see {\em e.g.} Ref.~\cite{Itzhaki}.

The intermediate region, $1/g_{YM}\sqrt{N_c}< x< \sqrt{N_c}/g_{YM}$,
where no perturbative field theoretical description
is possible, is fortunately exactly the region which is accessible to 
string theory; or rather, to the supergravity (SUGRA) approximation to 
Type IIB string theory on a special background.
It is that of the near horizon geometry of a
D1-brane in the string frame, which has the metric
\begin{eqnarray}\label{metric}
ds^2& =& \alpha' \hat{g}_{YM}
\left( {U^3 \over g^2_s}dx_\parallel^2 +  
{dU^2\over U^3} +  U d \Omega_{8-p}^2 \right) \nonumber \\
e^\phi & = & {2 \pi  g_{YM}^2 \over U^3}\hat{g}_{YM}.
\end{eqnarray}
where we defined $\hat{g}_{YM}\equiv 8\pi^{3/2} g_{YM}\sqrt{N_c}$.
In the description of the computation of 
the two-point function we follow Ref.~\cite{Anton98b}.  
The correlator has been derived in Ref.~\cite{ItzhakiHashimoto}, being itself
a generalization of Refs.~\cite{Gubser,Witten}.  

First, we need to know the action
of the diagonal fluctuations around this background to the quadratic
order. We would like to use the analogue of Ref.~\cite{KRvN} 
for our background, Eq.~(\ref{metric}), 
which is not (yet) available in the literature.
However, we can identify some diagonal fluctuating degrees of freedom
by following the work on black hole absorption
cross-sections \cite{krasnitz1,krasnitz2}. One can show
that the fluctuations parameterized like
\begin{eqnarray}
ds^2 & = & \left(1 +  f(x^0,U) +  g(x^0,U) \right) g_{00} (dx^0)^2 \nonumber\\
&& +
\left(1 +5 f(x^0,U) +  g(x^0,U)\right) g_{11} (dx^1)^2  \nonumber \\
&& + \left(1 +  f(x^0,U) +  g(x^0,U)\right) g_{UU} dU^2 \nonumber\\
&&+ \left(1 +
f(x^0,U) -    {5 \over 7} g(x^0,U)\right) g_{\Omega\Omega} d \Omega_7^2
\nonumber \\
e^\phi &=& \left(1 + 3 f(x^0,U) - g(x^0,U) \right) e^{\phi_0},
\end{eqnarray}
satisfy the following equations of motion
\begin{eqnarray}\label{fgeq}
f''(U)&=&-{7 \over U}  f'(U) + {g^2_s k^2 \over U^{6}} f(U)
  \\
g''(U) &=&   -{7 \over U} g'(U)+ {72 \over U^2} g(U)  + {g^2_s
k^2 \over U^6} g(U). \nonumber  
\end{eqnarray}
Without loss of generality we have assumed here that these
fluctuation vary only along the $x^0$ direction of the world volume
coordinates, and behave like a plane wave. One can interpret a D1-brane 
as a black hole in nine dimensions. The fields $f(U)$ and
$g(U)$ are then the minimal and the fixed scalars in
this black hole geometry. In ten dimensions, however, we see that they
are really part of the gravitational fluctuation. Consequently, we expect 
that they are associated with the stress-energy tensor in the operator
field correspondence of Refs.~\cite{Gubser,Witten}. In the case of the
correspondence between ${\cal N}{=}4$ SYM field theory and string 
theory on an $AdS_5 \times S^5$ background, the
superconformal symmetry allows for the identification of operators and
fields in short multiplets \cite{ferrara}. In the present case of a
D1-brane, we do not have superconformal invariance and this technique is not
applicable. Actually, we expect all fields of the theory consistent
with the symmetry of a given operator to mix.  The large distance
behavior should then be dominated by the contribution with the longest
range. The field $f(k^0,U)$ appears to be the one with the longest
range since it is the lightest field.

Eq.~(\ref{fgeq}) for $f(U)$ can be solved explicitly 
\begin{equation}
f(U) = U^{-3}  K_{3/2} \left( {\hat{g}_{YM}\over 2 U^{2}} k  \right),
\end{equation}
where $ K_{3/2}(x)$ is a modified Bessel function.
If we take  $f(U)$ to be the analogue of the minimally coupled
scalar, we can construct the flux factor
\begin{eqnarray} 
{\cal F} &=& \lim_{U_0 \rightarrow \infty} {1 \over 2
\kappa_{10}^2} \sqrt{g} g^{UU} e^{-2 (\phi - \phi_{\infty})}\nonumber\\
&& \left.\quad\quad\quad\quad\quad\quad\quad\quad\times \partial_U
\log( f(U))  \right|_{U = U_0} \nonumber\\
&=& {N U_0^2 k^2\over 2 g_{YM}^2} - {N^{3/2}
k^3 \over 4 g_{YM}} + \ldots
\end{eqnarray}
up to a numerical coefficient of order one which we have suppressed.
We see that the leading non-analytic  contribution in $k^2$ is due to
the $k^3$ term.  Fourier transforming the latter yields
\begin{equation}
\langle {\cal O}(x) {\cal O} (0) \rangle = {N^{{3 \over 2}} \over
g_{YM} x^5}. \label{SG}
\end{equation}
This is in line with the discussion at the beginning of this section.
We expect to deviate from the trivial ($1/x^4$)
scaling behavior of the correlator at $x_1={1}/{g_{YM}\sqrt{N_c}}$ and 
$x_2={\sqrt{N_c}}/{g_{YM}}$.
This yields the phase diagram in Fig.~\ref{figphase}.
It is interesting to note that
the entire $N_c$ hierarchy is consistent in the sense of Zamolodchikov's
c-theorem, which assures that the central charges obey 
$c(x)>c(y)$, whenever $x<y$ \cite{Zamo}.

\section{The correlator from SDLCQ}

Discretized Light-Cone Quantization (DLCQ) 
preserves supersymmetry at every stage
of the calculation if the supercharge rather than the Hamiltonian is 
diagonalized \cite{Sakai95}. 
The framework of supersymmetric DLCQ (SDLCQ) 
allows one to use the advantages of 
light-cone quantization ({\em e.g.~}a simpler vacuum) 
together with the excellent renormalization properties guaranteed by
supersymmetry.
Using SDLCQ, we can reproduce the SUGRA scaling relation, Eq.~(\ref{SG}), 
fix the numerical coefficient, and calculate the 
cross-over behavior at $1/g_{YM}\sqrt{N_c}<x<\sqrt{N_c}/g_{YM}$.
To exclude subtleties, {\em nota bene} issues of zero modes, 
we checked  our results 
against the free fermion and the 't Hooft model and found consistent
results.

The technique of (S)DLCQ was reviewed in Ref.~\cite{BPP},
so we can be brief here.  The basic idea of light-cone quantization
is to parameterize space-time using light-cone coordinates 
\begin{equation}
x^\pm\equiv \frac{1}{\sqrt{2}}\left(x^0\pm x^1\right),
\end{equation}
and to quantize the theory making $x^+$ play the role of time.
In the discrete light-cone approach, we require the momentum $p_- =
p^+$ along the $x^-$ direction to take on discrete values in units of
$p^+/K$ where $p^+$ is the conserved total momentum of the system.
The integer $K$ is the so-called harmonic resolution, and plays the role 
of a discretization parameter.
One can think of this discretization as a consequence of compactifying
the $x^-$ coordinate on a circle with a period $2L = {2 \pi K /
p^+}$. The advantage of discretizing on the light cone is the fact that
the dimension of the Hilbert space becomes finite.  Therefore, the
Hamiltonian is a finite-dimensional matrix, and its dynamics can be
solved explicitly.  In SDLCQ one makes the DLCQ approximation to the
supercharges $Q^i$. Surprisingly, also the discrete representations of $Q_i$
satisfy 
the supersymmetry algebra. Therefore SDLCQ enjoys the improved
renormalization properties of supersymmetric theories.  To
recover the continuum result, $K$ has to go to infinity.
Incidentally, what one finds is that SDLCQ usually converges faster than the
naive DLCQ towards the continuum limit. 

Let us now return to the problem at hand. We would like to compute a
general expression for the correlator of the form 
$F(x^-,x^+) = \langle {\cal O}(x^-,x^+) {\cal O} (0,0) \rangle$\@.
In DLCQ one fixes the total momentum in the $x^-$ direction, and it is
natural to compute the Fourier transform and express it
in a spectrally decomposed form
\begin{eqnarray}
\tilde{F}(P_-,x^+) &=& {1 \over 2 L} \langle {\cal O}(P_-,x^+) {\cal O}(-P_-,
0) \rangle\nonumber \\
&=&\sum_n {1 \over 2 L} \langle 0| {\cal O}(P_-) | n
\rangle e^{-i P_+^n x^+} \nonumber\\
&&\quad\quad\times\langle n|  {\cal O}(-P_-,0) |0 \rangle\ .
\end{eqnarray}
The form of the correlation function in position space is 
then recovered by Fourier
transforming with respect to $P_- =K\pi/L$.  We can continue to Euclidean
space by taking $r = \sqrt{2 x^+ x^-}$ to be real. The result for the
correlator of the  stress-energy tensor is 
\begin{eqnarray}
F(x^-,x^+)&=&
\sum_n\left|{L \over \pi} \langle n | T^{++}(-K) |0 \rangle \right|^2
\left({x^+ \over x^-}\right)^2\nonumber\\
&&\times {M_n^4 \over 8 \pi^2 K^3}
K_4\left(M_n\sqrt{2 x^+ x^-}\right),\label{master}
\end{eqnarray}
where $M_i$ is a mass eigenvalue and $K_4(x)$ is the
modified Bessel function of order 4. 
Note that this quantity depends on the harmonic resolution $K$, 
but involves no other unphysical quantities. In particular, 
the expression is independent of the box length $L$.

The momentum operator $T^{++}(x)$ of two-dimensional ${\cal N}=8$ SYM 
is given by
\begin{eqnarray}
T^{++} (x)&=&  {\rm tr} \left[ (\partial_- X^I)^2 \right.\nonumber\\
&&+\left. {1 \over 2} \left(i
u^\alpha \partial_- u^\alpha  - i  (\partial_- u^\alpha) u^\alpha
\right)\right], 
\label{susycorr}
\end{eqnarray}
with $I, \alpha = 1 \ldots 8$. 
$X$ and $u$ are the physical adjoint scalars and fermions,
respectively \cite{Anton98}.  When
discretized, these operators have the mode expansions
\begin{eqnarray}
X_{i,j}^I & = & {1 \over \sqrt{4 \pi}} \sum_{n=1}^{\infty} {1 \over
\sqrt{n}} \\
&&\left[ a^I_{ij} (n) e^{-i \pi n x^-/L} +
a^{\dagger I}_{ji}(n) e^{i \pi n x^-/L} \right],\nonumber \\
u_{i,j}^\alpha & = & {1 \over \sqrt{4 L}} \sum_{n=1}^{\infty} \\
&&\left[
b^\alpha_{ij} (n) e^{-i \pi n x^-/L} + b^{\dagger\alpha}_{ji}(-n) 
e^{i \pi n x^-/L}
\right].\nonumber
\end{eqnarray}
The matrix element $(L/\pi) \langle 0 | T^{++}(K) | i \rangle$ 
can be substituted directly to give an
explicit expression for the two-point function. We see immediately that
the correlator has the correct small-$r$ behavior, for in that limit, it
asymptotes to
\[
\left({x^- \over x^+ }\right)^2 F(x^-,x^+) =
   {N_c^2(2 n_b + n_f) \over 4  \pi^2 r^4}
            \left(1 - {1 \over K}\right),
\]
which we expect for the theory of $n_b(n_f)$ free bosons (fermions)
at large $K$.

On the other hand, the contribution to the correlator from strictly massless
states is given by
\begin{eqnarray}
&&\left( x^- \over x^+ \right)^2 F(x^-,x^+)=    {6 \over K^3 \pi^2 r^4}\\
   &&  \quad  \quad  \quad  \quad  \quad  \quad  
       \times \sum_i\left| {L \over \pi} 
            \langle 0 | T^{++}(K) | i\rangle \right|^2_{M_i=0}. \nonumber
\end{eqnarray}
It is important to notice 
that this $1/r^4$ behavior at large $r$ is {\em not} the one 
we are looking for at large $r$. First of all, we do not
expect any massless physical bound state in this theory, and,
additionally, it has the wrong $N_c$ dependence.
Relative to the $1/r^4$ behavior at small $r$, the $1/r^4$ behavior at
large $r$ that we expect is down by a factor of $1/N_c$. This behavior 
is suppressed because we are
performing a large-$N_c$ calculation. All we can
hope is to see the transition from the $1/r^4$ behavior at small $r$ to the
region where the correlator behaves like $1/r^5$.

\section{Symmetries and Numerics}

In principle, we can now 
calculate the correlator numerically by evaluating Eq.~(\ref{master}).
However, it turns out that even for very modest harmonic resolutions, 
we face a tremendous numerical task. 
At $K=2,3,4$, the dimension of the associated Fock space is
$256, 1632,$ and $29056$, respectively.
The usual procedure is to
diagonalize the Hamiltonian $P^-$ and then to evaluate the projection 
of each eigenfunction on the fundamental state $T^{++}(-K)|0\rangle$. 
Since we are only interested in states which
have nonzero value of such projection, we are able to significantly 
reduced our numerical efforts.

In the continuum limit, the result does not depend on which of the eight 
supercharges $Q^-_{\alpha}$ one
chooses. In DLCQ, however,  
the situation is a bit subtler: 
while the spectrum of $(Q_\alpha^-)^2$ is the same for all
$\alpha$, the wave functions depend on the choice of supercharge 
\cite{Anton98}. This dependence is an
artifact of the discretization and disappears in the continuum limit. 
What happens if we just pick one supercharge, say $Q^-_1$?
Since the state $T^{++}(-K)|0\rangle$
is a singlet under R--symmetry acting on the ``flavor'' index of
$Q^-_{\alpha}$, the correlator (\ref{master}) does not depend on the choice of
$\alpha$ even at finite resolution!

We can exploit this fact to simplify our calculations.
Consider 
an operator $S$ commuting with both $P^-$ and $T^{++}(-K)$, and such that
$S|0 \rangle=s_0|0 \rangle$. Then the Hamiltonian and $S$ can be diagonalized
simultaneously.  We assume in the sequel that the set of states
$|i \rangle$ is a result of
such a diagonalization. In this case, only states satisfying the condition
$S|i \rangle=s_0|i \rangle$ contribute to the sum in (\ref{master}), and we
only need to diagonalize $P^-$ in this sector, which reduces 
the size of the problem immensely.
We can deduce from the structure of the state
$T^{++}(-K)|0\rangle$ that any transformation of the form
\begin{eqnarray}
a^I_{ij}(k)\rightarrow f(I)a^{P[I]}_{ij}(k), \qquad f(I)=\pm 1
 \nonumber  \\ 
b^\alpha_{ij}(k)\rightarrow g(\alpha)b^{Q[\alpha]}_{ij}(k), \qquad
g(\alpha)=\pm 1 
\label{initsymm}
\end{eqnarray} 
given arbitrary permutations $P$ and $Q$ of the $8$ flavor indices, 
commutes with $T^{++}(-K)$.
The vacuum will then be an eigenstate of
this transformation with eigenvalue $1$. The requirement for $P^-=(Q^-_1)^2$
to be invariant under $S$ imposes some restrictions on the permutations.
In fact, we will require that $Q^-_1$ be invariant under $S$, in order
to guarantee that $P^-$ is invariant.

The form of the supercharge from \cite{Anton98} is
\begin{eqnarray}\label{Qminus}
&&Q^-_{\alpha} =  \int_0^{\infty}
  [...]b^{\dagger}_{\alpha}(k_3)a_{I}(k_1)a_{I}(k_2) +...\\
&&\!\!\!\!\!\!\!\!+(\beta_I \beta_J^T - \beta_J \beta_I^T )_{\alpha \beta}
  [..]  b^{\dagger}_{\beta}(k_3)a_{I}(k_1)a_{J}(k_2) + \ldots .\nonumber
\end{eqnarray}
Here the $\beta_I$ are $8\times 8$ real matrices satisfying
$\{\beta_I,\beta_J^T \} = 2\delta_{IJ}$. We use the special representation for
these matrices given in Ref.~\cite{GSW}.

Let us consider the expression for $Q^-_1$, Eq.~(\ref{Qminus}). 
The first part of the supercharge
does not include $\beta$ matrices, and is therefore invariant under
the transformation, Eq.~(\ref{initsymm}), 
as long as $g(1)=1$ and $Q[1]=1$. We will consider only such
transformations.
The crucial observation for the analysis
of the symmetries of the $\beta$ terms is 
that in the representation of the $\beta$ matrices we have chosen, 
the expression ${\cal B}^\alpha_{IJ}=
\left(\beta_I\beta^T_J-\beta_J\beta^T_I\right)_{1\alpha} $
may take only the values $\pm 2$ or zero. Besides, for any pair 
$(I,J)$ there is only one (or no) 
value of $\alpha$ corresponding to nonzero ${\cal B}$. Using this information,
we may represent ${\cal B}$ in a compact form. With 
the definition \cite{this work}
{\small
\begin{equation}
\mu_{IJ}=\left\{\begin{array}{rl} \alpha\,, & {\cal B}^\alpha_{IJ}=2 \\
                                  -\alpha\,, & {\cal B}^\alpha_{IJ}=-2 \\
                                     0\,,    & {\cal B}^\alpha_{IJ}=0
                                                {\mbox{ for all }}\alpha
                 \end{array}\right.,
\end{equation}
}
together with the special choice of $\beta$ matrices we get
the following expression for $\mu$
{\small
\[
\mu=\left(
\begin{array}{rrrrrrrr}
0&5&-7&2&-6&3&-4&8\\
-5&0&-3&6&2&-7&8&4\\
7&3&0&-8&-4&-5&6&2\\
-2&-6&8&0&-5&4&3&7\\
6&-2&4&5&0&-8&-7&3\\
-3&7&5&-4&8&0&-2&6\\
4&-8&-6&-3&7&2&0&5\\
-8&-4&-2&-7&-3&-6&-5&0\\
\end{array}
\right).
\]
}
The next step is to look for a subset of the transformations, 
Eq.~(\ref{initsymm}), which satisfy
the conditions $g(1)=1$ and $Q[1]=1$ and leave the matrix $\mu$ invariant. 
This invariance implies that
\begin{equation}
Q[\mu_{P[I]P[J]}]=g(\mu_{IJ})f(I)f(J)\mu_{IJ}\,.
\label{SymmFinalCond}
\end{equation}
The subset of transformations we are looking for forms a subgroup $R$ of
the permutation group $S_8\times S_8$. Consequently, we will search
for the elements of $R$ that
square to one. Products of such elements generate the whole group
in the case of $S_8\times S_8$.
We will show later that this remains true for $R$.
Not all of the $Z_2$ symmetries satisfying (\ref{SymmFinalCond})
are independent. In particular, if $a$ and $b$ are two such
symmetries then $aba$ is also a valid $Z_2$ symmetry. By going systematically 
through the different
possibilities, we have found that there are $7$ independent $Z_2$ symmetries in
the group $R$. They are listed in Table \ref{tab1}.
We explicitly constructed all the symmetries of the type, Eq.~(\ref{initsymm}),
which satisfy Eq.~(\ref{SymmFinalCond}) using {\sc Mathematica}. 
It turns out that the group 
of such transformations has $168$ elements, and we have shown that all
of them can be generated from the seven $Z_2$ symmetries mentioned above.
{\small
\begin{table*}
\centerline{
\begin{tabular}{|c|c|c|c|c|c|c|c|c|c|c|c|c|c|c|c|c|}
\hline
   &$a_1$&$a_2$&$a_3$&$a_4$&$a_5$&$a_6$&$a_7$&$a_8$&$b_2$&$b_3$&$b_4$&$b_5$&
   $b_6$&$b_7$&$b_8$\\
\hline
1 &$a_7$&$a_3$&$a_2$&$a_6$&$a_8$&$a_4$&$a_1$&$a_5$&$b_2$&$-b_3$&$-b_4$&$-b_6$&
   $-b_5$&$b_8$&$b_7$\\
\hline
2  &$a_3$&$a_6$&$a_1$&$a_5$&$a_4$&$a_2$&$a_8$&$a_7$&$-b_4$&$b_3$&$-b_2$&$-b_5$&
   $b_8$&$-b_7$&$b_6$\\
\hline
3  &$a_8$&$a_7$&$a_6$&$a_5$&$a_4$&$a_3$&$a_2$&$a_1$&$-b_3$&$-b_2$&$b_4$&$-b_5$&
   $b_7$&$b_6$&$-b_8$\\
\hline
4  &$a_5$&$a_4$&$a_8$&$a_2$&$a_1$&$a_7$&$a_6$&$a_3$&$-b_2$&$-b_7$&$b_8$&$b_5$&
   $-b_6$&$-b_3$&$b_4$\\
\hline
5  &$a_8$&$a_3$&$a_2$&$a_7$&$a_6$&$a_5$&$a_4$&$a_1$&$-b_5$&$-b_3$&$b_7$&$-b_2$&
   $b_6$&$b_4$&$-b_8$\\
\hline
6  &$a_5$&$a_8$&$a_7$&$a_6$&$a_1$&$a_4$&$a_3$&$a_2$&$-b_8$&$b_5$&$-b_4$&$b_3$&
   $-b_6$&$b_7$&$-b_2$\\
\hline
7  &$a_4$&$a_6$&$a_8$&$a_1$&$a_7$&$a_2$&$a_5$&$a_3$&$-b_2$&$-b_6$&$b_5$&$b_4$&
   $-b_3$&$-b_7$&$b_8$\\
\hline
\end{tabular}}
\caption{Seven independent $Z_2$ 
symmetries of the group $R$, 
which act on the 'flavor' quantum number of the different particles.
Under the first of these symmetries, {\em e.g.}, the boson $a_1$ is transformed
into $a_7$, etc.}
\label{tab1}
\end{table*}}

In our numerical algorithm we implemented the $Z_2$ symmetries as follows.
We can group the Fock states in classes and treat the whole class 
as a new state, because all states relevant for the correlator are 
singlets under the symmetry group $R$.
as an example consider the simplest non-trivial singlet 
\begin{equation}
|1\rangle=\frac{1}{8}\sum_{I=1}^8 {\mbox{tr}}
\left(a^\dagger(1,I)a^\dagger(K-1,I)\right)|0\rangle.
\end{equation}
Hence, if we encounter the state
$a^\dagger(1,1)a^\dagger(K-1,1)|0\rangle$ 
while constructing the basis, we will replace it by the class
representative; in this case, by the state $|1\rangle$. Such a procedure
significantly decreases the size of the basis, while keeping all the 
information necessary for calculating the correlator.
In summary, this use of the  
discrete flavor symmetry of the problem reduces the size 
of the Fock space by orders of magnitude.

In addition to these simplifications, one can further improve on the numerical 
efficiency by using Lanczos diagonalization techniques. 
Namely, we substitute the explicit diagonalization with an efficient 
approximation.
The idea is to use a symmetry preserving (Lanczos) algorithm.
If we start with a normalized vector $|u_1\rangle$ proportional to 
the fundamental state $T^{++}(-K)|0\rangle$,
the Lanczos recursion will produce a 
tridiagonal representation of the Hamiltonian
$H_{LC}=2P^+P^-$. 
Due to orthogonality of $\{|u_i\rangle\}$, 
only the (1,1) element of the tridiagonal matrix, $\hat{H}_{1,1}$, will 
contribute to the correlator.
We exponentiate by diagonalizing $\hat{H}_{LC}\vec{v}_i=\lambda_i\vec{v}_i$ 
with eigenvalues $\lambda_i$ and get 
\[
F(P^+,x^+)=\frac{|N_0|^{-2}}{2L}\left(\frac{\pi}{L}\right)^2
\sum_{j=1}^{N_L}|(v_j)_1|^2e^{-i\frac{\lambda_j L}{2K\pi}x^+}.
\] 
Finally, we Fourier transform to obtain
\begin{eqnarray}
F(x^-,x^+)&=&\frac{1}{8\pi^2K^3}\left(\frac{x^+}{x^-}\right)^2\frac{1}{|N_0|^2}\\
&&\times\sum_{j=1}^{N_L}|(v_j)_1|^2 \lambda_j^2 K_4(\sqrt{2x^+x^-\lambda_i}),
\nonumber
\end{eqnarray}
which is equivalent to Eq.~(\ref{master}).
This algorithm is correct 
only if the number of Lanczos iterations $N_L$ runs up to the rank of 
of original matrix. But {\em in praxi}  
already a basis of about 20 vectors covers all leading contribution 
to correlator \cite{Haydock}.

\begin{figure}
\centerline{
\psfig{file=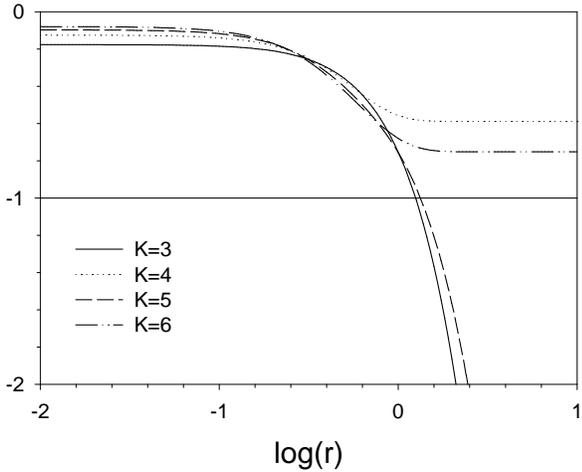,width=3.2in}} 
\centerline{(a)}
\centerline{
\psfig{file=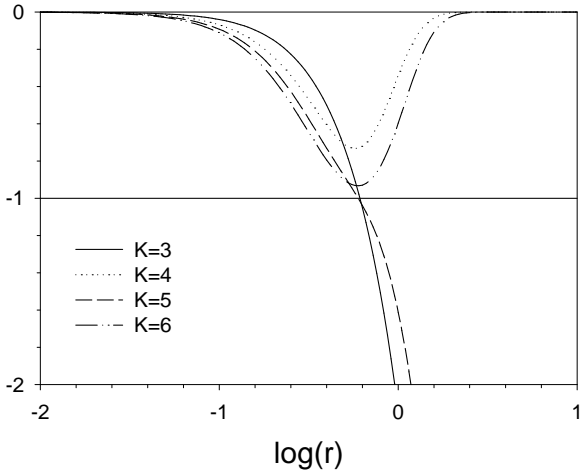,width=3.2in}} 
\centerline{(b)}
\caption{Top: (a) Log-Log
plot of ${\cal F}(r)=\langle T^{++}(x) T^{++}(0) \rangle
\left({x^- \over x^+} \right)^2 {4 \pi^2 r^4 \over N_c^2 (2 n_b +n_f)}$
vs. $r$ for $g_{YM}^2 N_c /\pi = 1.0$, $K=3,4,5$ and
$6$. Bottom: (b) the log-log
derivative with respect to $r$ of the correlation function in (a).
\label{fig}}
\end{figure}


\section{Results}

To evaluate the expression for the correlator
${\cal F}(r)$, we have to calculate the mass spectrum and insert it into 
Eq.~(\ref{master}).
In the ${\cal N}=(8,8)$ supersymmetric Yang-Mills
theory the contribution of massless states becomes a problem.
These states exist in the SDLCQ calculation, but are unphysical.
It has be shown that theses states are not normalizable
and that the number of partons in these states is even (odd) for 
$K$ even (odd) \cite{Anton98}.
Because the correlator is only sensitive to two particle contributions,
the curves ${\cal F}(r)$ are different for even and odd $K$.
Unfortunately, the 
unphysical states yield also the typical $1/r^4$ behavior, 
but have a wrong $N_c$ dependence.
The regular $1/r^4$ contribution is down by $1/N_c$, so we cannot 
see this contribution at large $r$, because we are working in the large 
$N_c$ limit.

We can use this information about the unphysical states, however, 
to determine when our approximation breaks down.
It is the region where the unphysical massless states dominate the
correlator sum.  Unfortunately, this is also the region where we
expect the true large-$r$ behavior to dominate the correlator, if
only the extra states were absent.
In Fig.~\ref{fig}(a) for even resolution, the region where the
correlator  starts to behave like
$1/r^4$ at large $r$
is clearly visible. In Fig.~\ref{fig}(b) we see that for even
resolution the effect of the massless state on the derivative
is felt at smaller values
of $r$ where the even resolution curves start to turn up. 
Another estimate of where this
approximation breaks down, that gives consistent values, is the set of
points where the even and odd resolution derivative curves cross. We do
not expect these curves to cross on general grounds, based on work
in \cite{Anton98b}, where we considered a number of other theories.
Our calculation is consistent in the sense that this breakdown 
occurs at larger and larger $r$ as $K$ grows.

We expect to approach the line $d{{\cal F}(r)}/dr=-1$ line 
signaling the cross-over from the trivial $1/r^4$ behavior 
to the characteristic $1/r^5$ behavior of the supergravity correlator, 
Eq.~(\ref{SG}). Indeed, the derivative curves in Fig.~\ref{fig}(b)
are approaching $-1$ as we increase the
resolution and appear to be about $85-90\%$ of this value 
before the approximation
breaks down. There is, however, no indication of convergence yet; therefore,
we cannot claim a numerical proof of the Maldacena conjecture.
A safe signature of equivalence of the field and string theories 
would be if the derivative curve 
would flatten out at $-1$ before the approximation breaks down.


\section{Conclusions}

In this note we reported on progress in an attempt to rigorously test
the conjectured equivalence of two-dimensional ${\cal N}=(8,8)$ 
supersymmetric Yang-Mills theory and a
system of $D1$-branes in string theory. 
Within a well-defined non-perturbative calculation, 
we obtained results that are within 10-15\% of results 
expected from the Maldacena conjecture. 
The results are still not conclusive, but they definitely 
point in right direction.
Compared to previous work \cite{Anton98b}, we included orders of magnitude
more states in our calculation and 
thus greatly improved the testing conditions.
We remark that improvements of the code and the numerical method 
are possible and under way.
During the calculation we noticed that contributions 
to the correlator come from only a small number of terms.
An analytic understanding of this phenomenon 
would greatly accelerate calculations.
We point out that in principle we 
could study the proper $1/r$ behavior at large $r$ by 
computing $1/N_c$ corrections, but this interesting calculation 
would mean a huge numerical effort.  

\begin{figure}
\centerline{
\psfig{file=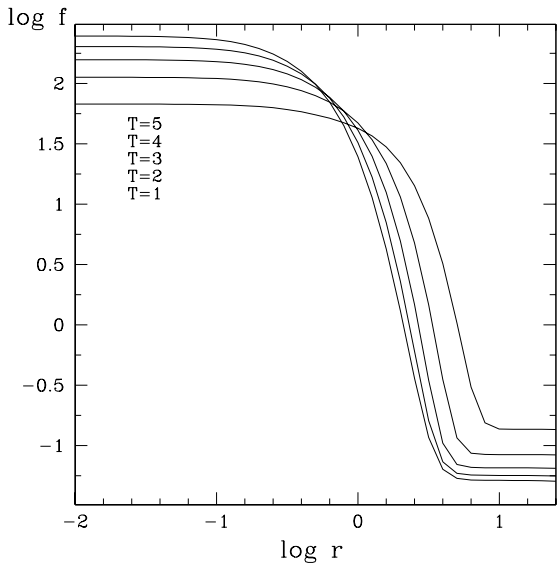,width=3in}} 
\label{plotnow}
\vspace*{-0.8cm}
\caption{Log-Log
plot of the {\em three-dimensional} correlation function $f\equiv r^5
\langle T^{++}(x) T^{++}(0) \rangle
\left({x^- \over x^+} \right)^2 {16\pi^3 K^3 l\over 105\sqrt{-i}}$
vs. $r$ for $g=g_{YM}^2 \sqrt{N_c l} /2\pi^{3/2} = 1.0$ for $K=6$ and
$T=1$ to $5$.}
\vspace*{-0.5cm}
\end{figure}


\section{Outlook}

It remains a challenge to rigorously test the conjectured
string/field theory correspondences. Although the so-called Maldacena 
conjecture maybe the most exciting one, because it promises insight 
into full four-dimensional Yang-Mills theories in the strong coupling regime,
there are other interesting scenarios. For instance, it was conjectured that
the supergravity solutions corresponding to $p+1$ SYM theories are black 
$p$-brane solutions, see {\em e.g.} Ref.~\cite{Itzhaki}. 
Consequently, there are interesting testing scenarios also in 
three-dimensional spacetime. Numerically, of course, things get the more 
difficult, the more dimensions are involved. 
On the way to the full four-dimensional problem, it may be worthwhile to 
present 
our latest results on correlation functions in three dimensions, see
also \cite{now}. Fig.~\ref{plotnow} shows the correlator for 
${\cal N}=1$ SYM(2+1) as a function of the distance $r$: it is 
converging well with the {\em transverse} cut-off $T$.
To put things in perspective, we note that the largest Hamiltonian 
matrix involved in this calculations requires to set up a 
Fock basis of approximately two million states. This is by a factor 100 more
than we used in the test of the Maldacena conjecture described in this 
article, which itself was already substantially 
better than the first feasibility study \cite{Anton98b}.

We hope to proceed on this way and to be able to present interesting
results soon.

\section*{Acknowledgements}

It is a pleasure to thank the organizers for the opportunity to speak
at this 'historic' workshop and for the hospitality during the symposium.
The excellent working conditions during the workshop are gratefully 
acknowledged,
which allowed for discussions leading to the project, Ref.~\cite{now}.

\end{document}